\documentclass[%
 aip,
 amsmath,amssymb,
 reprint,%
]{revtex4-1}

\usepackage{graphicx}
\usepackage{dcolumn}
\usepackage{bm}

\usepackage[utf8]{inputenc}
\usepackage[T1]{fontenc}
\usepackage{mathptmx}
\usepackage{etoolbox}
\usepackage[colorlinks=true,
                linkcolor=blue,
	        citecolor=blue,
	        urlcolor=blue]{hyperref}
\usepackage{soul}

\makeatletter
\def\@email#1#2{%
 \endgroup
 \patchcmd{\titleblock@produce}
  {\frontmatter@RRAPformat}
  {\frontmatter@RRAPformat{\produce@RRAP{*#1\href{mailto:#2}{#2}}}\frontmatter@RRAPformat}
  {}{}
}%

\makeatother
\begin{document}

\preprint{AIP/123-QED}

\title[]{Defects, Corrugation and Temperature Govern Rarefied-Air Drag on Graphene Coatings}

\author{Samuel Cajahuaringa}
\affiliation{Dipartimento di Fisica, Universit\`a di Trieste, I-34151 Trieste, Italy}
\author{Davide Bidoggia}
\affiliation{Dipartimento di Fisica, Universit\`a di Trieste, I-34151 Trieste, Italy}
\author{Maria Peressi}
\affiliation{Dipartimento di Fisica, Universit\`a di Trieste, I-34151 Trieste, Italy}

\author{Antimo Marrazzo}
\affiliation{%
Scuola Internazionale Superiore di Studi Avanzati (SISSA), I-34136 Trieste, Italy
}

\date{\today}

\begin{abstract}
In rarefied atmospheric environments, where continuum fluid dynamics breaks down, aerodynamic drag is governed by gas–surface momentum exchange, making surface structure and chemistry key design knobs. Using molecular dynamics simulations, we show that coating the $\alpha$-Al$_2$O$_3$(0001) surface with graphene markedly reduces the tangential momentum accommodation coefficient (TMAC) of N$_2$, shifting scattering toward more specular reflection and thereby lowering drag; we further benchmark this response against graphite. The reduction strengthens up to 900 K. While structural defects can increase TMAC via defect-induced corrugation and local atomic and electronic rearrangements, graphene retains its performance at experimentally relevant defect densities. 
%
\end{abstract}

\maketitle

The ability to tailor surface drag properties is of profound technological importance, particularly in aerospace applications where vehicles operate in rarefied atmospheric regimes. These environments are characterized by Knudsen numbers exceeding unity, when conventional continuum fluid dynamics fails and gas-surface interactions (GSI) become the dominant mechanism governing momentum and energy exchange~\cite{bird1994,boyd2017,sharipov2016}. Under such conditions, the drag force is no longer a bulk fluid property but is dictated by atomistic processes at the interface, making surface engineering a critical strategy for enhancing performance, energy efficiency, and thermal management. 

Graphene could be a promising coating for rarefied-flow applications because its atomic smoothness, high in-plane stiffness, and relative chemical inertness can promote more specular molecular reflections and reduce tangential momentum transfer to the wall. It is therefore interesting not only as a robust coating, but also as a model surface for understanding gas--surface scattering in the non-continuum regime. Tribological studies support this picture: graphene exhibits exceptional mechanical strength, chemical stability, and ultralow solid--solid friction, yet these properties are highly sensitive to lattice perfection~\cite{fan2024structure,ayyagari2022progress,jin2023lubrication,jafari2024tribological,Zambudio2021,banhart_structural_2011}.

In rarefied flows, drag is governed by the tangential momentum accommodation coefficient (TMAC), which measures the efficiency of tangential momentum transfer during molecular collisions with a surface. Lower TMAC values correspond to more specular, slip-like reflections and are therefore favorable for drag reduction, especially for high-altitude and hypersonic platforms~\cite{lee2022}. Although graphene is well known for reducing solid--solid friction, its influence on gas--surface drag remains largely unexplored.

A key question is how robust this low-accommodation behavior remains under realistic conditions, where graphene is rarely defect-free. Defects such as Stone--Wales (SW) transformations and vacancies disrupt the hexagonal lattice, generate local strain, and induce out-of-plane corrugation, thereby modifying the interaction landscape sampled by impinging gas molecules. As shown in the defect-engineering literature~\cite{bhatt_various_2022,tiwari_stonewales_2023}, even sparse defects can strongly alter interfacial response. This suggests that, just as defects tune graphene's tribological behavior, they may also affect gas-surface momentum accommodation and, ultimately, rarefied-flow drag.

In this work, we investigate how coating alumina with graphene\textemdash particularly, graphene containing realistic crystalline defects—modifies the drag experienced by nitrogen, the primary constituent of air, in rarefied environments. We compute TMAC and other gas\textemdash surface accommodation coefficients as functions of temperature and defect concentration, thereby quantifying how variations in defect density and thermal conditions jointly control rarefied drag, also unraveling the critical role played by defect-driven corrugation. 

Aluminum and its alloys are among the most widely used structural materials in aerospace systems~\cite{aglawe2023,li2023}. However, upon exposure to ambient air, aluminum rapidly forms a thin native oxide layer composed primarily of amorphous alumina (Al$_2$O$_3$)~\cite{cabmott1949,davis1993,evertson2015}.
This naturally passivating oxide layer, rather than the bare metal, constitutes the true interface with the surrounding gas. Consequently, we consider Al$_2$O$_3$ crystals to be a physically realistic proxy for aluminum alloys in studies of air drag under rarefied conditions.

\begin{figure*}[tbp]
    \centering
        \includegraphics[width=1.0\linewidth]{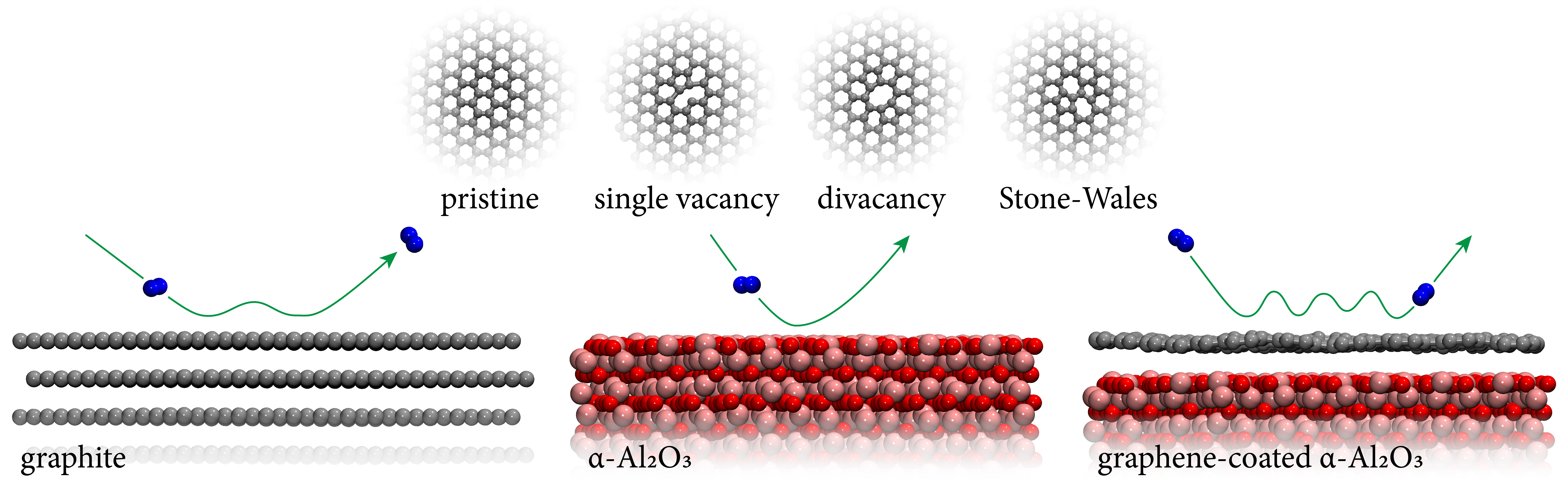}         
    \caption{Schematic representation of the simulated systems. Nitrogen molecule impinging on the multilayer graphite surface, the $\alpha$-Al$_2$O$_3$ (0001) slab, and the graphene-coated $\alpha$-Al$_2$O$_3$ (0001) surface. The inset shows the pristine graphene lattice and the optimized atomic configurations of the key defects studied: single vacancy, divacancy, and Stone–Wales defect.}
    \label{fig:surfaces}
\end{figure*}
We focus on three different surfaces: graphite, $\alpha-$Al$_{2}$O$_{3}$ (0001), and graphene-coated $\alpha-$Al$_{2}$O$_{3}$ (see Fig. \ref{fig:surfaces}). We use Monte Carlo (MC) sampling of thousands of molecular dynamics (MD) trajectories, which are obtained using the open-source LAMMPS~\cite{lammps} package and HyperQueue~\cite{HyperQueue} (more details in the Supplementary Material (SM)).

The graphite surface is modeled as a multilayered slab comprising 1690 carbon atoms arranged in five stacked honeycomb layers (left slab in Fig. \ref{fig:surfaces}). The alumina substrate is represented by a stoichiometric Al-terminated $\alpha-$Al$_{2}$O$_{3}$ (0001) slab—the thermodynamically most stable termination due to its lowest surface energy among all possible $\alpha-$Al$_{2}$O$_{3}$  facets~\cite{hutner_stoichiometric_2024,marmier_ab_2004,Ou2022,Ramogayana2021,Biaggne2021}. The $\alpha-$Al$_{2}$O$_{3}$ slab contains 36 atomic layers, i.e., 12 repetitions of the bulk unit triple-layer motif: Al–O$_{3}$–Al (central slab in Fig. \ref{fig:surfaces}). Details on the generation and thermalization of the slabs from NPT and NVT MD are discussed in the SM.

For graphene-coated alumina (right slab in Fig. \ref{fig:surfaces}), we take into account the inherent lattice mismatch between the hexagonal unit cells of graphene and $\alpha-$Al$_{2}$O$_{3}$ (0001). This mismatch induces interfacial strain~\cite{Acharya_gr_alumina_big_ACSanm2025}, which promotes spontaneous wrinkling and corrugation of the graphene layer~\cite{thiemann_defect-dependent_2021,leyssale2014}. To mitigate artificial strain and achieve a more physically realistic interface, we identify a common supercell that minimizes the strain on graphene cell by rotating graphene with respect to alumina~\cite{lazic_2015,carnevali2021}. Details about the interface generation are discussed in Sec.~S2 and Fig.~S2 of SM.
%
%
Nitrogen molecules (N$_{2}$) are initially positioned approximately 12 \AA~ above the topmost surface layer of each slab (Fig.~\ref{fig:surfaces}). This distance exceeds the cutoff radius of the interatomic potential that models the GSI, ensuring that molecules begin their trajectories in free space, initially unaffected by van der Waals (vdW) attraction. This setup allows for a clean initialization of the collision event, consistent with the methodology employed in prior MD studies of gas-surface scattering~\cite{andric_2018,thoudam_2024}. Interatomic potentials are selected based on the specific material system: For the graphite slab, the AIREBO~\cite{stuart_2000} potential is used to describe C–C bonding and long-range dispersion interactions. For the $\alpha-$Al$_{2}$O$_{3}$ slab, the many-body Vashishta~\cite{vashishta_2008} potential is employed to accurately capture the complex ionic and covalent interactions within the oxide lattice. Because graphene adsorbs only weakly on alumina~\cite{Acharya_gr_alumina_big_ACSanm2025} the graphene-alumina interaction is modeled by a pairwise Lennard-Jones (LJ) potential parameterized to reproduce the DFT-derived adsorption curve~\cite{Storn1997}, computed using \textsc{Quantum ESPRESSO} ~\cite{Giannozzi_2009, Giannozzi_2017, Giannozzi2020} with the Perdew-Burke-Ernzerhof (PBE) functional~\cite{Perdew_PRL_1996} plus Grimme’s D3 dispersion correction~\cite{grimme_jcp_2010} and using SSSP Precision v1.3 pseudopotentials~\cite{prandini_npjcm_2018,dhv_prb_1990,pslib0.3,pslib1} (see Sec.~S1 of the SM for more details).

\begin{figure*}[tbp]
    \centering
    \begin{tabular}{cccc}
        \includegraphics[width=0.26\linewidth]{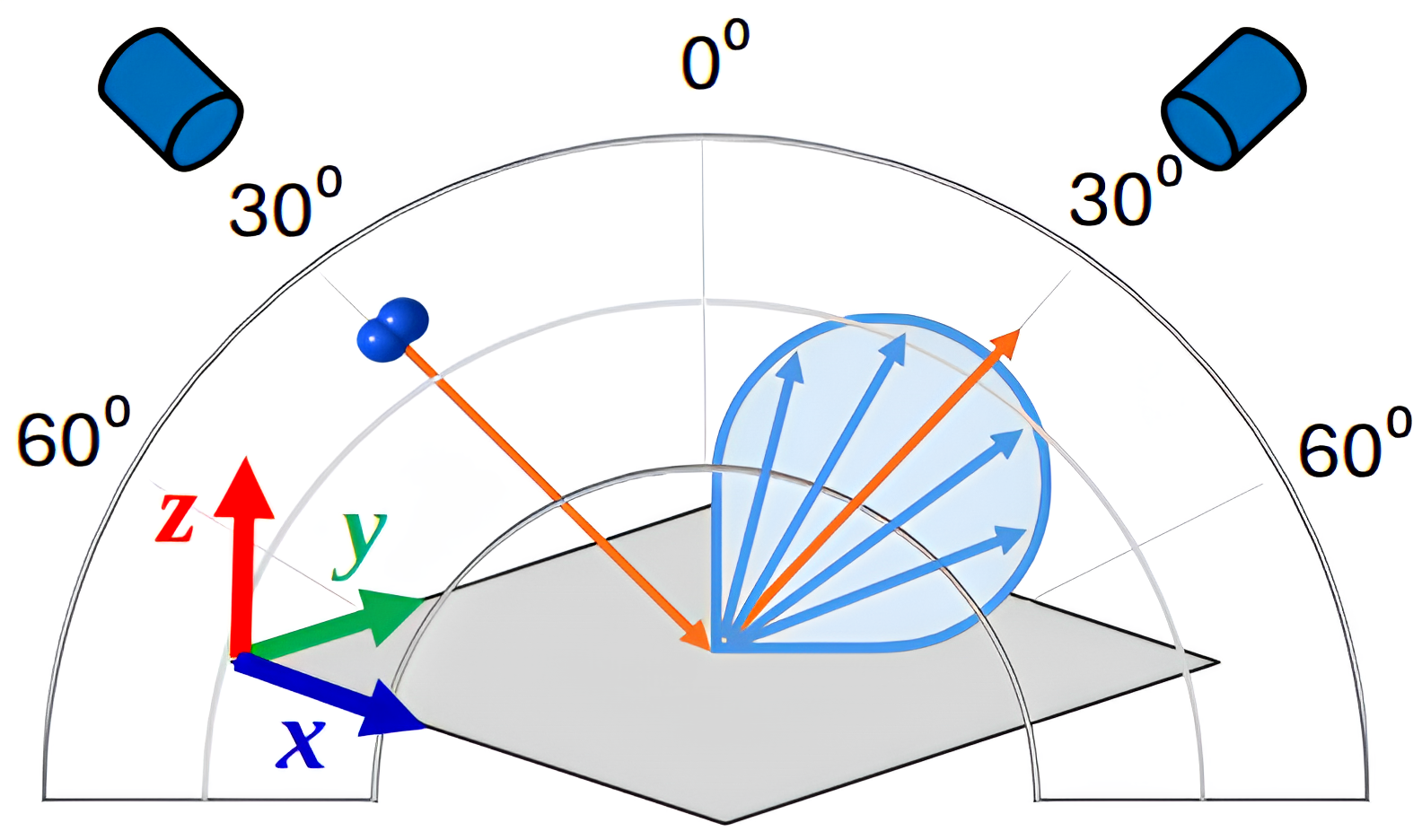} &
        \includegraphics[width=0.20\linewidth]{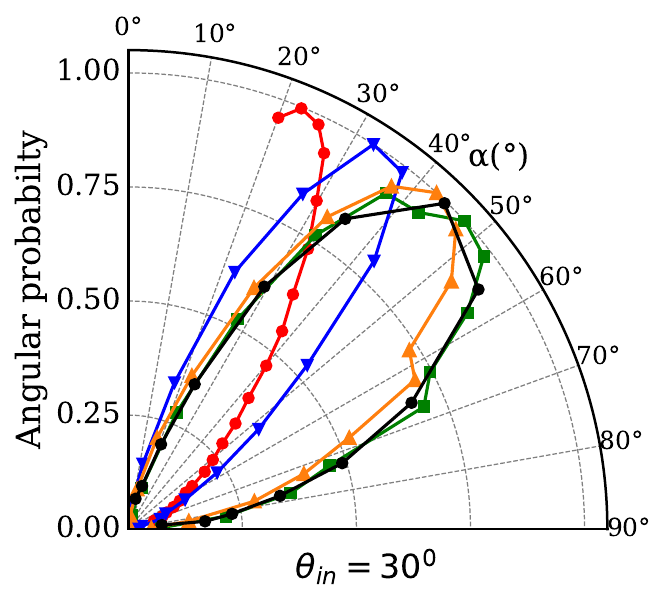} &
        \includegraphics[width=0.20\linewidth]{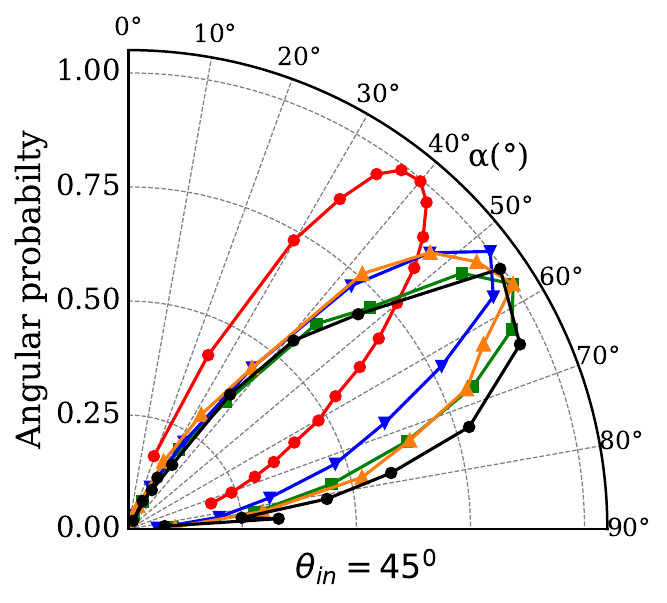} &
        \includegraphics[width=0.26\linewidth]{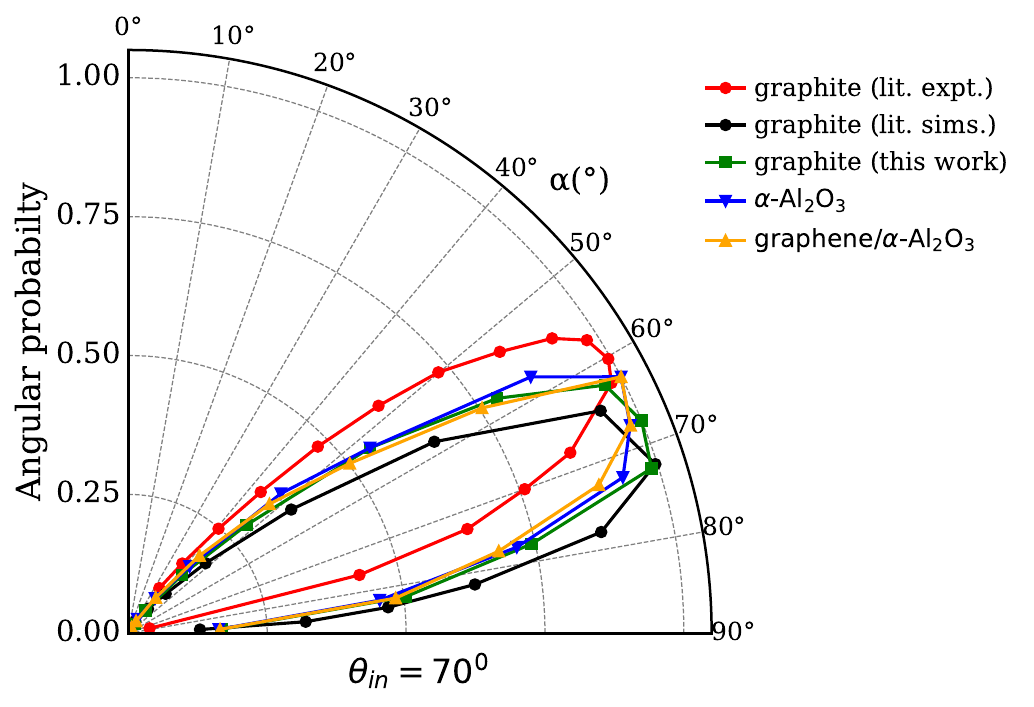} \\
        a & b & c & d \\
    \end{tabular}
    \caption{(a) Schematic representation of gas beam scattering at fixed angle. (b-d) Angular distribution of scattered nitrogen molecules on selected surfaces as a function of reflection angle $\alpha$ for three different incidence angles (b) $\theta_{in}=30^{\circ}$; (c) $\theta_{in}=45^{\circ}$; (d) $\theta_{in}=70^{\circ}$.  Our results for $\mathrm{graphite}$, $\alpha$-$\mathrm{Al_{2}O_{3}}$, $\mathrm{graphene}$/$\alpha$-$\mathrm{Al_{2}O_{3}}$ are shown as green line with circles, blue line with squares, orange line with triangles, respectively.  Experimental~\cite{mehta_2018,andric_2018} and simulation~\cite{andric_2018} results on $\mathrm{graphite}$ from the literature are marked with red lines with circles and black lines with circles, respectively. A version of these plots in Cartesian coordinates is available in the SM as Fig. S6.}
    \label{fig:angular_distribution}
\end{figure*}

As N$_{2}$ is the dominant constituent of air (78\% by mole fraction), it is a good proxy to study the effect of air drag; the inclusion of oxygen and water is left to future studies. N$_{2}$ is modeled as a rigid diatomic molecule and vibrational degrees of freedom are frozen using the SHAKE~\cite{Shake} algorithm to constrain the bond length at its equilibrium value (1.0976 \AA). This approximation is well justified for the temperature range considered in this work is below 2000 K and nitrogen molecules remain in the vibrational ground state, while rotational-translational energy exchange dominates the GSI  dynamics. The intermolecular potentials governing GSI are selected based on the substrate: for N$_{2}$ interacting with graphite or graphene, the LJ potential is employed, as validated in prior molecular beam and MD studies of N$_{2}$–carbon systems~\cite{yamanishi_1999, andric_2018}, while for N$_{2}$ interacting with $\alpha-$Al$_{2}$O$_{3}$ (0001) we use the Morse potential, consistent with recent simulations of N$_{2}$–alumina scattering in noncontinuum regimes~\cite{thoudam_2024}. 
A key aspect of this work is the use of two complementary MD protocols targeting distinct physical regimes of gas–surface collisions. The first protocol mimics molecular gas beam experiments, in which a collimated beam of N$_{2}$ molecules impinges on the surface at fixed incidence angles and velocities (Fig.~\ref{fig:angular_distribution} and Fig. S3 in SM).
We perform molecular gas beam scattering simulations for nitrogen molecules impinging on graphite, bare $\alpha-$Al$_{2}$O$_{3}$ (0001), and graphene-coated $\alpha-$Al$_{2}$O$_{3}$ (0001) surfaces. This enables us to validate our simulation framework, particularly the accuracy of GSI potentials, against established experimental and theoretical benchmarks for graphite\textemdash a system extensively studied in both beam experiments and MD studies~\cite{andric_2018, mehta_2018, mehta_2017}. While prior studies have reported discrepancies between some numerical predictions and experimental angular distributions for graphite, our primary objective here is not to achieve perfect quantitative agreement with experiment. Instead, we aim to elucidate qualitative behavior and relative trends in scattering across different surface types, with particular emphasis on isolating the respective impacts of structural defects and temperature.

To this end, we replicate the well-characterized experimental setup of Metha et al.~\cite{mehta_2018}: a supersonic N$_{2}$ beam with a nominal velocity of 1453 m/s incident at angles of $30^{\circ}$, $45^{\circ}$, and $70^{\circ}$ onto a graphite surface held at 677 K under high-vacuum conditions~\cite{mehta_2017,andric_2018,mehta_2018}. The same configuration is applied to alumina-based surfaces to investigate how surface chemistry and graphene coating influence scattering dynamics. For each incidence angle, 4000 independent trajectories are simulated per surface, with a maximum simulation time of 100 ps per trajectory to prevent spurious trapping (details in Sec. S5 of the SM). Initial molecular positions, orientations, and velocities are sampled via Monte Carlo importance sampling \cite{cajahuaringa2023} (details in Sec. S2 of the SM), ensuring statistical representativeness and accurate reproduction of beam conditions.

Trajectories are terminated when the scattered molecule reaches a height exceeding the intermolecular cutoff distance above the surface, guaranteeing free-flight conditions for post-collision analysis. The final translational velocities are used to compute angular distributions as a function of the scattering angle $\alpha$, which reflect the probability of reflection at a given angle. Figures~\ref{fig:angular_distribution}b--d show the computed angular distributions for pristine alumina, graphene-coated alumina, and graphite at several incidence angles $\theta_{\mathrm{in}}$, together with experimental data and simulations from the literature for graphite. Our graphite results agree with prior computational studies~\cite{andric_2018} and qualitatively reproduce the characteristic scattering profiles reported in experiments~\cite{yamanishi_1999,mehta_2017}. Because the angular distribution governs the gas-surface momentum exchange, these features directly influence the aerodynamic drag.

At small incidence angles (Fig.~\ref{fig:angular_distribution}b,c), the angular distribution for graphene-coated alumina is closer to that of graphite than to that of bare alumina. At the largest incidence angle, $\theta_{\mathrm{in}}=70^{\circ}$ (Fig.~\ref{fig:angular_distribution}d), the distributions converge and all three surfaces become nearly indistinguishable.

This behavior can be rationalized as follows. As the angle of incidence increases (at fixed incident speed), the component of the kinetic energy normal to the surface decreases. The incident normal energy therefore becomes small, and energy exchange is dominated by coupling to the thermal motion of the surface atoms. In this regime, particles do not penetrate deeply into the surface potential and primarily sample an effectively averaged, weakly corrugated top layer. For graphite and graphene-coated alumina, this top layer can be viewed as a nearly monoatomic, almost flat surface, whereas for alumina it is a slightly corrugated diatomic surface due to the non perfect co-planarity of  Al and O atoms and their different vdW radii, 1.84~\AA{} and 1.52~\AA{}, respectively. At higher normal incident energies, by contrast, particles can probe and interact with the lateral corrugation of the surface potential, entering the structural scattering regime. Scattering then becomes more sensitive to surface topography and corrugation, and the angular distributions broaden. The relatively low normal energies associated with large incidence angles therefore lead to similar angular distributions for all three surfaces, consistent with previous observations~\cite{Rettner_1991,mehta_2017}.
%
%
\begin{figure*}[tbp]
    \centering
    \begin{tabular}{ccc}
        \includegraphics[width=0.32\linewidth]{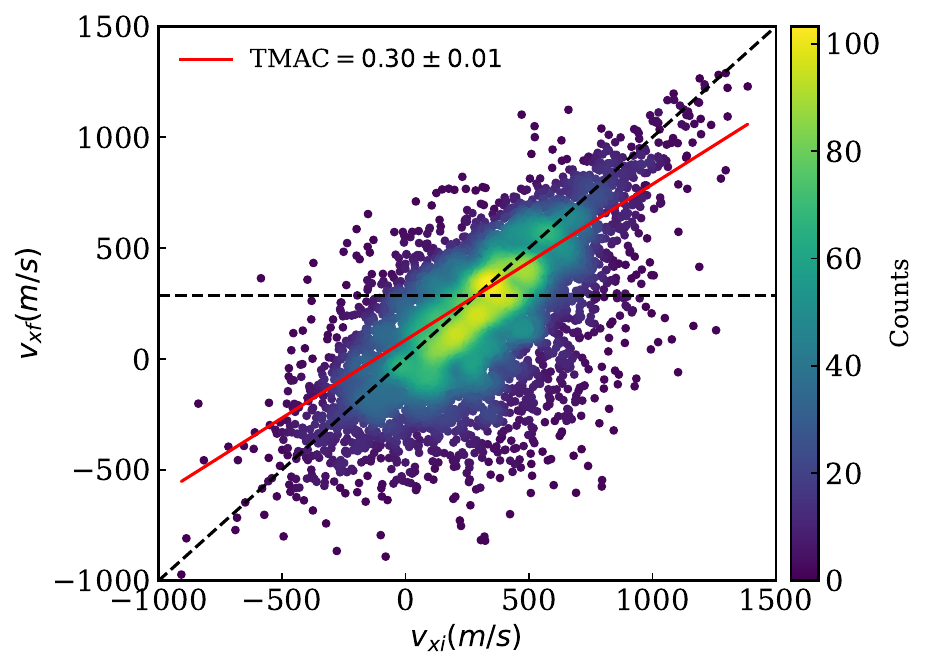} &
        \includegraphics[width=0.32\linewidth]{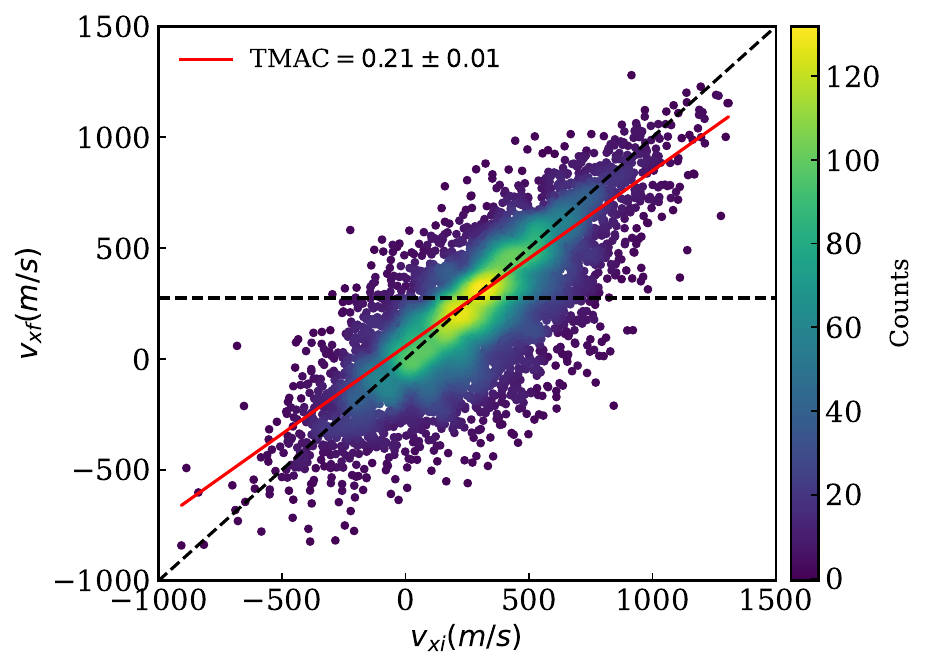} &
        \includegraphics[width=0.32\linewidth]{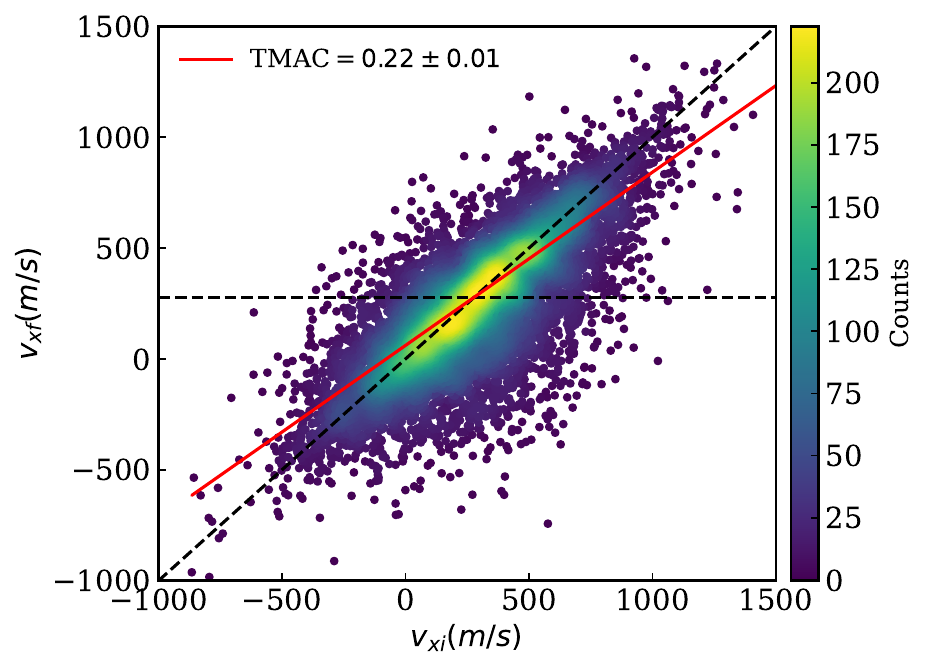} \\
        $\alpha-\mathrm{Al}_{2}\mathrm{O}_{3}$ & graphene/$\alpha-\mathrm{Al}_{2}\mathrm{O}_{3}$ & graphite \\
    \end{tabular}

    \caption{Correlations between incoming (horizontal-axis) $v_{xi}$ and outgoing final $v_{xf}$ (vertical-axis) tangential velocity of nitrogen molecules on three different surfaces: alumina, graphene-coated alumina and graphite, for a gas flow at room temperature with bulk velocity of 280 m/s. The solid red lines demonstrate the least-square linear fit of the tangential velocity data to obtain the tangential momentum accommodation coefficients (TMAC).}
    \label{fig:TAC_x}
\end{figure*}
These results, based on collimated molecular beams with prescribed incidence angle and speed, provide detailed insight into GSI and show that graphene-coated alumina behaves more similarly to graphite than to bare alumina. However, these beam-type simulations are conditional on a narrowly defined initial molecular state.
To overcome this limitation, we now move from fixed, collimated beams\textemdash chosen previously to enable a direct comparison with molecular-beam experiments\textemdash to a configuration more relevant for assessing coating performance under realistic flow conditions. In this complementary approach, initial molecular velocities are randomly assigned by sampling from the equilibrium distribution at a given gas temperature. Molecules thus impinge with a distribution of incidence angles and kinetic energies (details in Sec. S4 of the SM). The theoretical foundations of this stochastic initialization have been extensively discussed in the literature and will not be repeated here~\cite{thoudam_2024,andric_2018,yamanishi_1999,mehta_2017}. In many practical situations, the initial molecular state is not fixed as in a molecular-beam experiment but reflects a thermalized gas.

Accommodation coefficients (ACs) provide a compact and widely used measure of gas-surface scattering, quantifying the efficiency of momentum and energy transfer during collisions, which could be defined as%
\begin{equation}
	\alpha_{q} = \frac{\left\langle Q_{I} \right\rangle - \left\langle Q_{R} \right\rangle}{\left\langle Q_{I} \right\rangle - \left\langle Q_{T} \right\rangle}
\end{equation}
where the subscript $q$ denotes a specific kinematic property of the gas molecule, such as its momentum or kinetic energy in a given direction. The brackets denote that the average values for these quantities need to be computed, and the subscripts denote whether the average value is to be computed from the incoming (I) particles, the reflected (R) particles or from the thermal wall distribution (T). The principal limitation of this conventional definition is that it becomes numerically unstable when the gas temperature approaches the surface temperature, i.e., near thermal equilibrium~\cite{spijker_2010} ($\langle Q_I \rangle \approx \langle Q_T \rangle$). This issue is particularly relevant here because our goal is to quantify drag under conditions where the gas and the surface might be in thermal equilibrium.

To overcome this limitation, we adopt the velocity-correlation-based method introduced by Spijker et al.~\cite{spijker_2010,Spijker_2008}, which remains robust even under isothermal conditions~\cite{spijker_2010,bayer-buhr_determination_2022,mohammad_nejad_influence_2020,mohammad_nejad_modeling_2021}. In this approach, various ACs are obtained from the slope of a least-squares linear fit to the MD collisional data expressed in terms of appropriately defined velocity correlation functions:
\begin{equation}
    \alpha_{q} = 1 - \frac{\sum_i\left( Q_{I}^{i} - \left\langle Q_{I} \right\rangle \right)\left( Q_{R}^{i} - \left\langle Q_{R} \right\rangle \right)}{\sum_i\left( Q_{I}^{i} - \left\langle Q_{I} \right\rangle \right)^{2}},
\label{eq:ac_correlation}    
\end{equation}
where $Q^{i}_{I}$ and $Q^{i}_{R}$ represent, respectively, the pre-collisional and post-collisional values of the same property for the $i$-th gas particle. We define the TMAC for nitrogen flow along the x direction as:
\begin{equation}
    TMAC = 1 - \frac{\sum_i\left( p_{I,x}^{i} - \left\langle p_{I,x} \right\rangle \right)\left( p_{R,x}^{i} - \left\langle p_{R,x} \right\rangle \right)}{\sum_i\left( p_{I,x}^{i} - \left\langle p_{I,x} \right\rangle \right)^{2}}  
\end{equation}
We calculate TMACs for nitrogen flow at room temperature, with a bulk flow velocity of 280~m/s (representative of a commercial aircraft flight speed), over surfaces also held at room temperature. For each of the three surfaces considered, a total of 10\,000 trajectories is simulated to obtain statistically converged TMAC values. The resulting scattering data used to compute the TMACs are summarized in Fig.~\ref{fig:TAC_x}. Following the approach of Spijker et al.~\cite{spijker_2010}, the tangential velocities of molecules crossing a virtual plane before and after collision with the surface are recorded and used to determine the ACs via a linear least-squares fit.  Figure~\ref{fig:TAC_x} shows the correlation between incoming and outgoing tangential velocities employed to compute TMAC. The limiting cases TMAC $=0$ and TMAC $=1$ are indicated by black dashed lines: the diagonal line corresponds to fully correlated velocities, i.e., purely specular scattering with TMAC $=0$, whereas the horizontal line corresponds to completely uncorrelated velocities, i.e., fully accommodated particles with TMAC $=1$. The red line represents the best-fit linear regression to the MD data, and its slope is used to obtain the value of TMAC. An increase in the slope corresponds to a decrease in TMAC. For all surfaces, the incoming and outgoing tangential velocities are distributed around the diagonal line and are centered near 280~m/s, corresponding to the imposed flow velocity. For the bare alumina, graphene-coated alumina, and graphite surfaces, we obtain TMAC values of 0.30(1), 0.21(1), and 0.22(1), respectively. The value for alumina is consistent with that reported by Thoudam et al.~\cite{thoudam_2024}. Inspection of the correlation plots reveals that the fraction of specular-like collisions increases for the graphene-coated alumina surface. We observe an approximate 30\% reduction in the TMAC of alumina when coated with graphene, approaching that of graphite.

In the rarefied regime, aerodynamic drag is determined by the net momentum flux exchanged between incident and reflected molecules at the surface. Accordingly, a lower TMAC implies more specular-like reflection and reduced tangential momentum transfer to the wall, whereas a higher TMAC promotes more diffuse scattering and larger drag, consistent with classical free-molecular theory and with Maxwell/CLL-based kinetic and DSMC studies showing a strong sensitivity of drag coefficients to the adopted gas-surface interaction model and accommodation parameters \cite{sentman_lmsc_1961,Lord1992,walker_2014,Kalempa2020}. From this perspective, the TMAC is a microscopic descriptor of gas-surface scattering, whereas the drag coefficient is a macroscopic aerodynamic quantity resulting from the cumulative effect of many such collisions over the entire body. Therefore, while ACs more directly reflect the intrinsic material-dependent properties of the gas-surface interaction, the drag coefficient also depends on external flow conditions, such as gas composition, velocity, and temperature, as well as on the geometry and orientation of the object.

\begin{figure}[tbp]
    \centering
    \includegraphics[width=1.0\linewidth]{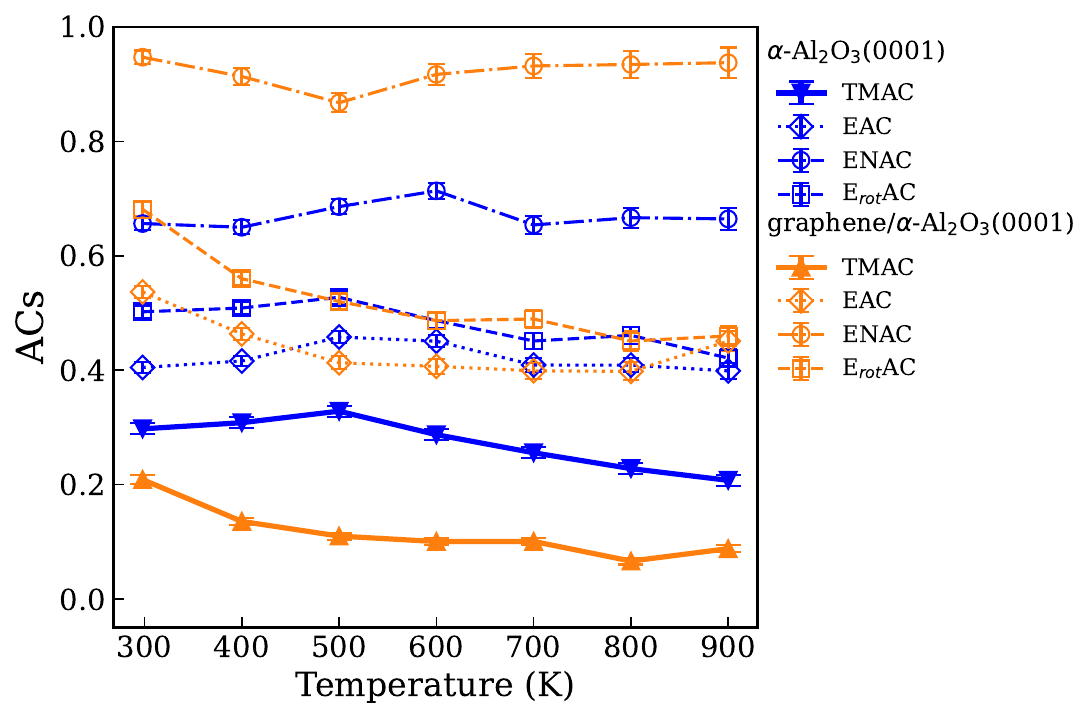}
    \caption{Temperature dependence of ACs: tangential momentum accommodation (TMAC), energy accommodation coefficient (EAC), normal energy accommodation coefficient (ENAC) and rotational energy accommodation coefficient (E$_{\mathrm{rot}}$AC) for bare alumina (blue lines) and graphene-coated alumina (orange lines). Error bars represent the standard error of the mean computed from the linear regression used to obtain the ACs.}
    \label{fig:ACs_vs_temperature}
\end{figure}
Studying the temperature dependence of ACs is essential for understanding GSI mechanisms under realistic thermal conditions. In particular, the TMAC directly influences heat and momentum transfer in applications such as thermal protection systems, micro- and nanoscale gas flows, and heterogeneous catalysis. Fig.~\ref{fig:ACs_vs_temperature} shows the variation of ACs as a function of temperature for two surfaces: bare alumina (blue curves) and graphene-coated alumina (orange curves). For alumina, TMAC exhibits a slight increase up to approximately 500~K, followed by a gradual decrease with further increase in temperature. In contrast, for the graphene-coated alumina surface, TMAC starts from a lower value and displays a more pronounced decay as temperature rises. This behavior indicates that the graphene layer significantly reduces gas-surface momentum exchange and enhances specular reflection, particularly at elevated temperatures. A similar trend has been reported for graphite in a hydrogen environment, where the TMAC decreases with increasing temperature and tends to level off at high temperatures~\cite{kovalev_2010}.

We also compute additional ACs to obtain a more comprehensive picture of the GSI dynamics (see Fig.~\ref{fig:ACs_vs_temperature}). For the energy accommodation coefficient (EAC), which measures the overall energy transfer from the gas to the surface, the bare alumina surface shows little sensitivity to temperature in range of 0.4 to 0.445, whereas its graphene-coated counterpart exhibits a clear decrease in EAC with increasing temperature. The normal energy accommodation coefficient (ENAC), quantifying the exchange of kinetic energy in the direction perpendicular to the surface, displays only a weak temperature dependence for both surfaces. The rotational energy accommodation coefficient (E$_{\mathrm{rot}}$AC) reveals the most pronounced difference between the two materials. For bare alumina, E$_{\mathrm{rot}}$AC decreases for temperatures above 500~K, mirroring the trend observed for the TMAC. A similar, but more marked, monotonic decrease with temperature is found for the graphene-coated surface. In summary, Fig.~\ref{fig:ACs_vs_temperature} demonstrates that graphene coating not only reduces TMAC but this reduction is also present in EAC and E$_{\mathrm{rot}}$AC for increasing temperatures (all correlations graphs for ACs at different temperatures are reported in Fig. S7 and S8 of the SM). 
\begin{figure}[tbp]
    \centering
    \includegraphics[width=0.8\linewidth]{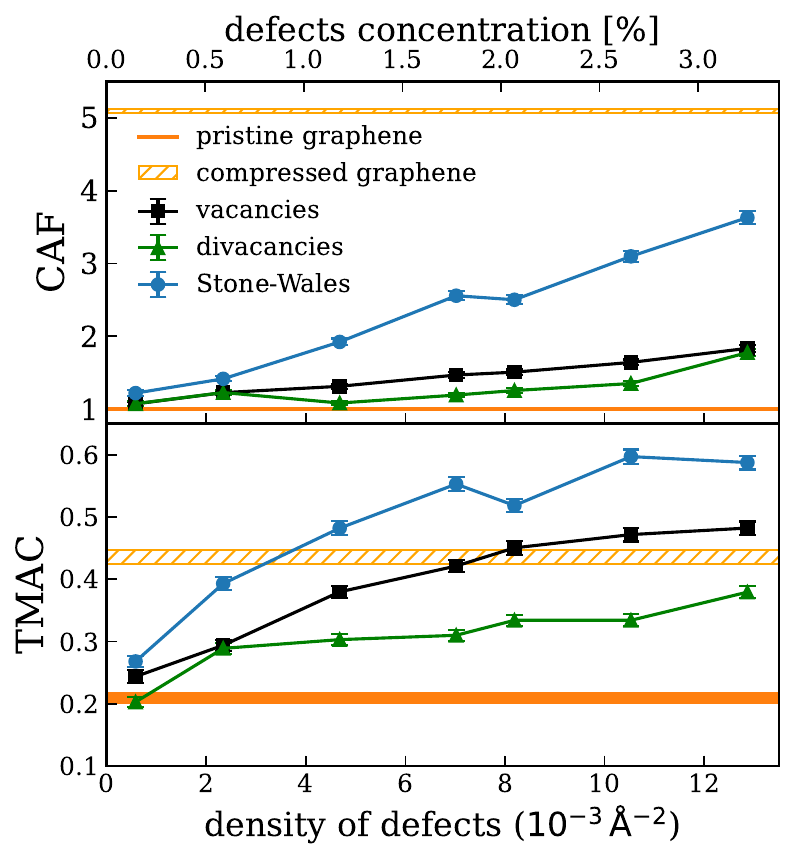}
    \caption{CAFs (top) and TMACs (bottom) and at different density of defects on graphene-coated alumina for vacancies (black squares), divacancies (green triangles) and Stone-Wales (light blue circles). The orange and light-orange dashed lines correspond respectively to pristine and 2\% compressed graphene.}
    \label{fig:TAC_vs_concentration}
\end{figure}

Up to this point, our analysis has focused on the properties of pristine graphene-coated alumina, an idealized scenario. Under realistic conditions, however, graphene is rarely defect-free and typically contains structural imperfections such as vacancies, divacancies, and SW defects~\cite{banhart_structural_2011,bhatt_various_2022}. These defects can significantly modify the electronic, structural, and interfacial properties of graphene, and may therefore influence the underlying GSI mechanisms. Unlike generic rough surfaces, where drag is mainly governed by roughness-induced diffuse scattering, graphene combines exceptional atomic flatness, high mechanical stiffness, and weakly interacting carbon chemistry. These properties can favor more specular reflections and reduced momentum transfer in the pristine case; accordingly, vacancies and divacancies may enhance accommodation not only through local geometric disruption, but also through defect-driven ripples~\cite{thiemann_defect-dependent_2021, thiemann_pnas_2025} and possibly chemical interactions \cite{Carraro2022,Jovanovic2023, Kumar2020}. 

To address this, we investigate how the presence of vacancies, divacancies, and SW defects in graphene coating alumina affects the TMAC. By quantifying the impact of these defects, we aim to gain deeper insight into how realistic surface conditions govern momentum exchange and energy dissipation during gas-surface collisions. Fig.~\ref{fig:TAC_vs_concentration} shows that defects significantly affect the TMAC of graphene-coated alumina. Following Thiemann et al. \cite{thiemann_defect-dependent_2021}, we define the defect concentration as the ratio between the number of atoms removed (one/two per vacancy/divacancy) or displaced (two per SW defect) and the total number of atoms in a pristine graphene sheet. Defects are randomly distributed over the graphene layer, subject to the constraint that defect centers are separated by at least six neighbors from each other (inset in Fig.~\ref{fig:surfaces}). Fig.~\ref{fig:TAC_vs_concentration} shows the dependence of TMAC on defect density. For each defect density, a single representative configuration is considered; error bars reflect the statistical uncertainty associated with the finite sampling of MD collision trajectories and do not include configurational averaging. The results indicate that increasing the density of defects leads to a pronounced increase in TMAC. This trend is mirrored in Fig.~\ref{fig:TAC_vs_concentration} where the corrugation amplification factor (CAF) is calculated. Following Thiemann et al.~\cite{thiemann_defect-dependent_2021}, the CAF is defined as the ratio of the height-fluctuation standard deviation of a defective system to that of pristine graphene. As the defect density increases, the CAF rises, indicating that the surface becomes increasingly corrugated or ``wrinkled.'' This enhanced roughness disrupts the specular reflection mechanism characteristic of smooth surfaces and promotes more diffuse scattering events (more details in Sec. S7 of SM), in which tangential momentum is transferred more efficiently to the surface. The correlation between CAF and TMAC suggests that the enhanced surface roughness induced in particular by SW defects is the primary driver of increased tangential momentum transfer. In Sec. S8 of the SM, we report a sensitivity analysis that shows how CAF is robust to variations of the LJ interface well depth by $\pm10\%$.

To test whether corrugation alone can account for this effect, we simulate pristine graphene under 2\% compressive strain—a condition known to induce topographic undulations without introducing atomic-scale defects (details in  Sec. S9 of SM). Remarkably, the TMAC for compressed graphene matches that of vacancies-defective graphene at comparable (but lower) levels of CAF (Fig.~\ref{fig:TAC_vs_concentration}a, dashed orange line). In particular, a TMAC value attained by compressed graphene at $\mathrm{CAF} \gtrsim  5.0$ is reproduced in the presence of SW defects at substantially lower corrugation, with $\mathrm{CAF} < 2$. This suggests that, although corrugation appears to be the dominant effect for SW defects, it is unlikely to be the sole mechanism responsible for the associated increase in drag. This is much more evident for vacancies and divacancies, which also lead to significant TMAC enhancement (Fig.~\ref{fig:TAC_vs_concentration}), yet they produce far less corrugation compared to SW defects at equivalent defect densities (Fig.~\ref{fig:TAC_vs_concentration}). Despite lower CAF values, vacancies exhibit TMAC increases comparable to those seen in highly corrugated systems.

This decoupling of TMAC from CAF implies that surface roughness is not the dominant mechanism behind TMAC in case of vacancies and divacancies. Instead, the local lattice distortion around the defects to be the critical factor, de-correlate the trajectories of nitrogen molecules and promote more diffuse scattering. The distinct responses associated with different defect types indicate that multiple physical pathways can enhance TMAC: one mediated by geometric roughness (e.g., SW-driven or strain-induced corrugation), and another driven by topological disruption of the lattice (e.g., vacancies and divacancies).

Experimental studies have shown that the typical vacancy concentration in graphene at room temperature ranges from approximately $10^{-6}$ to $10^{-4}$~vacancies/\AA$^{2}$, depending on the synthesis route and irradiation conditions~\cite{thiemann_defect-dependent_2021,tiwari_stonewales_2023,bhatt_various_2022}. In our simulations, we explore defect concentrations in the range $5.8\times10^{-4}$ to $1.287\times10^{-2}$~vacancies/\AA$^{2}$. Lower, experimentally relevant defect densities were not simulated explicitly because they would require substantially larger surface cells and therefore prohibitively expensive trajectory sampling, while also making defect-hit events increasingly rare, so their accommodation coefficients are expected to lie even closer to the pristine-surface limit. Within this range, the lowest TMAC for the vacancy case is 0.24(1), compared with 0.21(1) for pristine graphene-coated alumina, corresponding to a modest increase of about 14\%. By contrast, the lowest TMAC for the divacancy case is 0.20(1), effectively identical to that of pristine graphene. These results suggest that, over the range of experimentally relevant defect concentrations, the TMAC is expected to remain essentially unchanged, and a typical density of vacancies in graphene is unlikely to significantly alter gas-surface momentum exchange. Moreover, the trends obtained at higher defect densities may provide useful insight into the behavior of graphene-based coatings under wear or harsh operating conditions, where damage accumulation can generate defect populations well above those expected for as-prepared graphene.

In conclusion, our results demonstrate that coating alumina with a graphene layer effectively reduces the TMAC and, consequently, the drag force at the gas--surface interface. This reduction indicates that the graphene layer promotes more specular gas reflection and limits momentum transfer to the substrate. The effect persists across the entire temperature range considered, with the gap between the TMAC of graphene-coated and bare alumina even increasing with temperature, showing that the low-drag characteristics of graphene are not only maintained but enhanced at high temperatures. A possible extension of the present framework would be to assess non-adiabatic energy-dissipation channels, including electronic excitations during gas-surface collisions, which could be relevant and interesting to explore under more extreme impact conditions. The presence of graphene defects substantially increases drag, an effect partially attributed to defect-induced corrugation. Nevertheless, within the range of experimentally relevant defect concentrations, the TMAC of graphene-coated alumina increases only modestly. Therefore, even under realistic conditions where defects are unavoidable, graphene-coated alumina is expected to retain its ability to significantly reduce the TMAC and thus the drag force at the gas--surface interface.

\vspace{-18pt}
\section*{Supplementary Material}
\vspace{-12pt}
The Supplementary Material contains more details on computational protocols and additional results for ACs and CAF.

\vspace{-18pt}
\begin{acknowledgments}
\vspace{-12pt}
All authors acknowledge useful discussions with A. Abdurrazaq, D. Acharya, S. de Gironcoli, C. Di Valentin, D. Dragoni, A. Maslov, Z. Muhammad. and C. Scalliet.  The work was supported from the ICSC -- Centro Nazionale di Ricerca in HPC, Big Data and Quantum Computing, funded by European Union - NextGenerationEU (CUP Grant. No. J93C22000540006, PNRR Investimento M4.C2.1.4), and Leonardo S.p.A. through the Innovation Grant ASGARD. The authors acknowledge CINECA, under CINECA-SISSA, CINECA-UniTS and ICSC agreements, for the availability of high-performance computing resources and support.  The views and opinions expressed are solely those of the authors and do not necessarily reflect those of the European Union, nor can the European Union be held responsible for them.
\end{acknowledgments}
\vspace{-12pt}
\section*{AUTHOR DECLARATIONS}
\vspace{-12pt}
\subsection*{Conflict of Interest}
\vspace{-12pt}
The authors have no conflicts to disclose.
\vspace{-20pt}
\subsection*{Author Contributions}
\vspace{-12pt}

\textbf{Samuel Cajahuaringa:} Data curation (lead); Investigation (lead); Methodology (lead); Resources (equal); Software (lead); Visualization (lead); Writing – original draft (lead); Writing – review \& editing (equal). \textbf{Davide Bidoggia:} Data curation (supporting); Investigation (supporting); Methodology (supporting);  Visualization (supporting); Writing – original draft (supporting); Writing – review \& editing (equal).  \textbf{Maria Peressi:} Funding acquisition (equal), Investigation (supporting); Methodology (supporting); Project administration (equal);  Supervision (equal); Visualization (supporting); Writing – review \& editing (equal). \textbf{Antimo Marrazzo:} Funding acquisition (equal), Conceptualization (lead); Investigation (supporting); Methodology (supporting); Project administration (equal); Resources (equal); Supervision (equal); Visualization (supporting); Writing – original draft (supporting); Writing – review \& editing (equal).
\vspace{-22pt}
\section*{Data Availability}
\vspace{-12pt}
The data that support the findings of this study are openly available on the Materials Cloud~\cite{Data_available}.
\vspace{-22pt}
\section*{References}
\vspace{-22pt}
\bibliographystyle{aipnum4-1_new}
\bibliography{biblio}

\end{document}